\documentstyle[prb,aps,epsf,amstex,a4]{revtex}
\begin{document}
\draft
\title{The Randomly Driven Ising Ferromagnet\\ Part II: One and two dimensions\\}
\author{Johannes Hausmann$^{\dagger}$ and P\'al Ruj\'an$^{\dagger,\$}$}
\address{Fachbereich 8 Physik$^{\dagger}$ and ICBM$^{\$}$, Postfach 2503, Carl
  von Ossietzky Universit\"at, D-26111 Oldenburg, Germany}
 
\date{\today} \maketitle
\begin{abstract}
  \noindent
  We consider the behavior of an Ising ferromagnet obeying the Glauber
  dynamics under the influence of a fast switching, random external
  field. In Part~I, we introduced a general formalism for
  describing such systems and presented the mean-field theory. In this
  article we derive results for the one dimensional case,
  which can be only partially solved. Monte Carlo
  simulations performed on a square lattice indicate that the main
  features of the mean field theory survive the presence of strong
  fluctuations.
\end{abstract}


This paper considers the randomly driven Ising model (RDIM), introduced
and discussed from a general point of view in Part I\cite{part1},
in one and two dimensions. The main interest in studying such 
highly nonlinear nonequilibrium statistical physical systems lies in
their possible applications for storage and other information processing 
tasks. Many biological systems, especially networks of real neurons,
share common features with the RDIM in that they form strongly
coupled systems driven by external stimuli acting in unison over a macroscopic
number of elements in characteristic times shorter than the typical
system relaxation time. Therefore, the Gaussian or Poissonian external
``noise'' is transformed radically within the coupled system, leading
to a ``correlated noise'' with a lot of ``strange'' properties. 
The mean field theory was developed in
Part I \cite{part1} for a random binary switching external field.
Applications to cortical neurons will be discussed elsewhere.
 
The order parameter of the RDIM is a nonequilibrium
stationary magnetization distribution, which undergoes a symmetry
breaking bifurcation at some critical field strength. The analytic
structure of this distribution also changes in character as a function
of the temperature and field parameters. Hence, transitions
between a
singular function with fractal support to a singular function with euclidean
support and further, to an absolutely continuous distribution can be observed.
They can be seen in finite size effects and in the variance (the fluctuations) of the
free energy. Similar transitions can be seen also in one dimension.
However, the critical lower dimensionality of the RDIM remains one. Although
the static critical exponent is a function of the field/temperature ratio,
the dynamic exponent $\Delta=2$ remains unchanged. Some simple arguments
will be given, supporting these results.

The situation is much more difficult in two-dimensions, where only a
Monte Carlo simulation approach was possible. The main difficulty is that
computing the stationary properties of the system requires a large population
of different trajectories. We found that it is more convenient to use
systems of moderate size than to rely on only few dynamic trajectories.
Even so, the simulations require a vast amount of computer resources.
We solved this problem by using the Siemens-Nixdorf neurocomputer SYNAPSE-1N110
for quite a different task than what it was conceived for. This machine is basically a
matrix-computer, allowing us to run many different systems in parallel.
Even so, we were able only to determine roughly the phase diagram itself:
we have no means at the moment for calculating the critical exponents. 

The paper is organized as following: in Section~\ref{sec:1d} we consider
the one dimensional case. Although it cannot be fully solved, many interesting
exact results can be derived. Section~\ref{sec:mcs} deals with the
Monte Carlo simulations we performed on finite square lattices. Many features
of the mean-field dynamics are shown to survive the strong
fluctuations characteristic to two dimensional systems. However, in
contrast to the mean-field approach, the two dimensional model
displays also an interesting spatial structure related to droplet
dynamics. Comparisons between mean-field theory and two-dimensional
results are systematically presented, including some preliminary
results for hysteresis. Finally, we discuss our results in
Section~\ref{sec:sum}.





\section{RDIM in one dimension}
\label{sec:1d}
\subsection{The Master Equation}

In his pioneering work, Glauber \cite{glauber} defined a stochastic
dynamics where only single spin flips are allowed and hence neither
the magnetization (the order parameter) nor the energy is preserved
(model A universality class, see \cite{hohhalp}). Although introduced
mainly for mathematical convenience, this dynamics is believed to
describe appropriately many Ising-like systems.

The energy of an Ising chain with periodic boundary conditions is
given by
\begin{equation}
        \label{enising}
         E = -J\sum_{i=1}^N s_i s_{i+1} + \mu_B B(t) \sum_{i=1}^N s_i
\end{equation}
Following the notation of Part I\cite{part1} we denote by
the vector $\vec \mu=(s_1,s_2,\dots,,-s_i,\dots,s_N)$ a given configuration
of spins and by $\vec \mu_i=(s_1,s_2,\dots,-s_i,\dots,s_N)$ the same
configuration but with the spin $s_i\to -s_i$ flipped. The external field is
sampled from $\rho(B) = {1\over 2} \delta(B - B_0) + {1\over 2} \delta(B + B_0)$
at time intervals of length $\tau_B$,
\begin{equation}
    B(t) = \mu_B B \rho(B) \sum_{n=0}^{\infty}\Theta(t-n \tau_B)\Theta \left((n+1)\tau_B - t\right)
    \label{h(t)}
\end{equation}
As explained in Part I, the Master Equation has the form
\begin{eqnarray}
        \dot P(\{s_i\};t) & = & - \hat {\cal L}_{B(t)} P(\{s_i\};t)  \cr 
        & = & \sum_i^N w(\vec \mu | \vec \mu_i) P(\vec \mu_i;t) - P(\vec \mu;t) \sum_i^N w(\vec \mu_i | \vec \mu)  
        \label{meq1d}
\end{eqnarray}
where $w(s_i) \equiv w(\vec \mu | \vec \mu_i)$ is the transition rate from a configuration $\{s_i\}$ into the state where only the $i$-th spin is flipped,
$\{-s_i\}$.  

Strictly speaking, the system we are going to consider here will never
reach the equilibrium Boltzmann distribution ${\rm e}^{-\beta E}$,
where $\beta = 1/k_BT$. Nevertheless, we require the detailed balance
condition to be fulfilled for a constant value of the external field.
        
The transition probability $w(s_i)$ is determined by the constraint of
detailed balance only up to a positive arbitrary function of the
neighboring spins, $f(s_{i-1},s_{i+1})$. Since in one dimension the
phase transition is at $T_c=0$, the choice of the transition
probability influences the analytic form of the critical singularities
\cite{pfp}. In what follows we will use the form
\begin{equation} 
        \label{trprob}
        w(s_i) = {1\over {2 \alpha}} [1 - s_i {\rm tanh}(K\sum_{j\in \langle i,j \rangle} s_j + H)]
\end{equation}
where $K=\beta J$, $H=\beta \mu_B B$, $\alpha$ sets the time constant,
and $\langle i,j \rangle$ denotes next neighbor pairs. Glauber
introduced this form for $H=0$ but used a slightly different one for
$H\neq 0$. In one dimension one has then
\begin{equation}
        \label{trprob1d}
        w(s_i)  = {1\over {2 \alpha}} [1 - s_i {\rm tanh}(K(s_{i-1}+s_{i+1}) + H)] 
\end{equation}

For $H=0$ the Liouville operator $\hat {\cal L}_{B}$ can be mapped
onto a free-fermion spin-chain Hamiltonian \cite{siggia}. This
explains why the equations for the averaged spin products, $\langle
\prod_{j\in \alpha} s_j \rangle $, decouple in subspaces which can be
diagonalized by appropriate Fourier transformations. As shown below,
this property is inherited by the first moments, $[\pi_{\alpha}(t)]$
of the stationary distribution ${\cal P}_s$.

\subsection{Magnetization and correlation functions}

Consider, for example, the time evolution of the local magnetization
\begin{equation}
        m_i(t) = \langle s_i \rangle_t = \sum_{\{s_i\}} s_i P(\{s_i\};t)
        \label{mitdef}
\end{equation}
which can be obtained from \eqref{meq1d} as
\begin{equation}
        \dot m_i = -{2 \over \alpha} \langle s_i w(s_i) \rangle ={1\over \alpha} \left[ - m_i(t) + \langle  {\rm tanh}(K(s_{i-1}+s_{i+1}) + H)
          \rangle_t \right] 
        \label{dotmi}
\end{equation}
In order to make the relationship to chaotic maps more evident, we use
now the `coarse grained' form\cite{part1} of \eqref{meq1d}.  Formally, this procedure
corresponds to a forward Euler discretization of \eqref{dotmi},
setting the time step equal to $\tau_B$, and measuring time in units
of $\alpha = \tau_B $:
\begin{eqnarray} 
        m_i(t+1) & = & \langle  {\rm tanh}(K(s_{i-1}+s_{i+1}) + H) \rangle_t \nonumber \\ 
        & = &a + \tilde \gamma{ m_{i-1}(t) + m_{i+1}(t) \over 2} +  b \langle s_{i-1}s_{i+1} \rangle_t
        \label{mimap}
\end{eqnarray}
where $\gamma = {\rm tanh}2K$, $h = {\rm tanh}H$, and
\begin{eqnarray}
        a & = & {h\over 2} ({1-\gamma^2\over 1- h^2\gamma^2} + 1) \label{adef} \\
        \tilde \gamma & = & \gamma {1- h^2 \over 1- h^2\gamma^2}  \label{tgammadef} \\
        b & = & {h\over 2} ({1-\gamma^2\over 1- h^2\gamma^2} - 1) \label{bdef}
\end{eqnarray}

\noindent Similarly, for the correlation function
\begin{equation}
        c_{i,j} = \langle s_i s_j \rangle
        \label{cijdef}
\end{equation}
one obtains
\begin{equation}
        \dot c_{i,j} = -{2 \over \alpha} \langle s_i s_j (w(s_i) + w(s_j)) \rangle_t
        \label{dotcij}
\end{equation}
leading to the map
\begin{equation}
        c_{i,j}(t+1) = {1\over 2} \langle s_j {\rm tanh}(K(s_{i-1}+s_{i+1}) + H) + 
        s_i {\rm tanh}(K(s_{j-1}+s_{j+1}) + H) \rangle_t
        \label{cijmap}
\end{equation}
where the time unit is now set to $\alpha = 2 \tau_B$. These
recursions can be written again in terms of the variables $a$, $b$ and
$\tilde \gamma$, Eqs. (\ref{adef}-\ref{bdef}).

For $H=0$ one has $\tilde \gamma = \gamma$ and, as shown by Glauber
\cite{glauber}, the slowest relaxation time equals the inverse of the
smallest eigenvalue of the magnetization subspace, Eq. (\ref{dotmi}).
This relaxation time diverges with the square of the static
correlation length (the dynamic critical exponent is $z=2$).

For the binary field distribution  of \eqref{h(t)} the map of the local magnetization
\eqref{mimap} has two branches:
\begin{equation}
        \label{mibimap}
    m_i(t+1) = \left\{ \begin{array}{lcl} 
                     a + \tilde \gamma { m_{i-1}(t)+m_{i+1}(t) \over 2} + b c_{i-1,i+1}(t)\ & {\rm with\ prob.} &\ {1\over 2} \\
                     & & \\
                    -a + \tilde \gamma { m_{i-1}(t)+m_{i+1}(t) \over 2} -b c_{i-1,i+1}(t)\ & {\rm with\ prob.} &\ {1\over 2} \\
                        \end{array} \right.
\end{equation}
where it is implicitly assumed that the time needed to switch the
field is negligible, $\tau_{switch} << \tau_B$.

From Eq. \eqref{mibimap} it is evident that the map for the local
magnetization couples to a two-spin correlation, which in turn couples
to higher order correlations, etc.  Hence, the full dynamic map lives
in a $2^N$ dimensional space, as stated in Part I\cite{part1}.

Nevertheless, some partial results can be obtained for the stationary
distribution. Define a `thermal' and a `dynamical' average,  $\langle \dots \rangle$ resp.
$\left[ \dots \right]$. In the stationary state one has $[\langle A(t+1) \rangle ] =
[\langle A(t) \rangle ]$ for any spin-function $A$.  The average of
the local magnetization obeys
\begin{equation}
        \label{mistat}
        [m_i] = {[\tilde \gamma]\over 2} ([m_{i-1}]+ [m_{i+1}])
\end{equation}
where (recall that $\gamma = \tanh 2K$, $h_0=\tanh H_0$)
\begin{equation}
        \label{tgamma}
        [\tilde \gamma] = \gamma {1 - h_0^2 \over 1 - h_0^2 \gamma^2}
\end{equation}

For the translation invariant magnetization $m = {1\over N} \sum_i
m_i$ and the two-spin correlation function $c_m = {1\over N} \sum_i
c_{i,i+m}$ one obtains equations formally similar to the
ones solved by Glauber \cite{glauber}.  From \eqref{mistat}, except
for $[\tilde \gamma] = 1$, the magnetization vanishes. The stationary
two-spin correlation function obeys
\begin{equation}
        [c_m] = {[\tilde \gamma] \over 2} ([c_{m-1}] + [c_{m+1}])
       \label{cm}
\end{equation}
which leads to
\begin{equation}
        [c_m] = \eta^{|m|}\ \ {\rm with} \ \eta = {1-\sqrt{1-[\tilde \gamma]^2} \over [\tilde \gamma] }
\end{equation}

In general, when expressing the Frobenius-Perron operator in the basis
formed by all moments of spin-products $[\pi_{\alpha}^q(t)]$, the
subspace of the {\it first} moments ($q=1$) is closed and can be
diagonalized by a Fourier transform.  The remaining part, however, is
intractable.  For example, the second moment of the magnetization
reads
\begin{equation}
        [m^2] = [a^2] + 2[ab] [c_2] + [{\tilde \gamma}^2] [m^2] + [b^2] [c_2^2]
        \label{secmom}
\end{equation}
and couples both to the first and to the second moment of the
translation invariant correlation functions, $[c_2]$ and $[c_2^2]$,
respectively.

\subsection{ The $T_c=0$ phase transition}

As expected, in the stationary state the odd moments of the
magnetization vanish at $T>0$.  By expanding $\tilde \gamma$ and
$\eta$ at low temperature one obtains after straightforward
calculations that close to $T_c=0$ the correlation length is in
leading order
\begin{equation}
  \label{xi1d}
  \xi= - {1\over \ln \eta} \sim \left\{ {\begin{array}{lcl} 
        {1 \over 2} {\rm e}^{2K-H_0}\ & \ {\rm if} \ & 2K > H_0 \\
        & & \\
        {1 \over 2} {1 \over H_0 - 2K}\ & \ {\rm if} \ & 2K < H_0 \\
      \end{array}} \right.
\end{equation}
where $H_0= \beta \mu_B B_0$ and $B_0$ is defined as in Eq.~\eqref{h(t)}.
Hence, in one dimension the RDIM has a critical line with continuously
changing singularities at $T_c=0$ for $\kappa \equiv H_0/2K \in
[0,1]$.  The slowest relaxation time corresponds to the magnetization
(order parameter) decay and can be computed as
\begin{equation}
\label{tau1d}
\tau_{sys}^{-1} = 2(1-[\tilde \gamma]) \sim \left\{ \begin{array}{lcl}   \xi^{-2}\ & \ {\rm if} \ & \kappa < 1 \\
                                                                        & & \\
                                                                        2 \ & \ {\rm if} \ & \kappa > 1 \\
                                      \end{array} \right.
\end{equation}

Consider first the case of a strong field, $\kappa > 1$.  The field
will align all spins in one iteration step, as evident from Eq.
\eqref{tau1d}.  Hence, the spins are almost always parallel to the
driving field.  Since $[\tilde \gamma]$ vanishes, there is no
spontaneously broken symmetry and $[m]=0$ due to the symmetry of the
field distribution, $\rho(B)=\rho(-B)$.

For fields smaller than the critical field $\kappa_c = 1$ one obtains
a true symmetry breaking ferromagnetic phase.  Interesting enough,
while the divergence of the correlation length decreases continuously
according to Eq. \eqref{xi1d}, the critical dynamic exponent remains
$z=2$ up to and including $\kappa \leq 1$.  A physical argument
explaining this result is presented below.

\subsection{ Kink dynamics at $T=0$}

Consider now the transition probability $w(s_i)$ at $T=0$, Eq. \eqref{trprob1d}, which is
a function of the three spins $s_i$, $s_{i-1}$, and $s_{i+1}$. 
Let us call an interface between two oppositely oriented spin domains a {\it kink}.
If $\kappa = H_0/2K = 0$ and $T=0$ each kink performs
a random walk,  moving with equal probability to the left or to the right. 
When two kinks become nearest neighbors, 
they annihilate  
because in the next time step
the single spin left between them will flip with probability one.

Due to this annihilation process the number of 
kinks decreases steadily and in the end only very few are left.
Two kinks situated at the typical distance $\xi$ (the correlation length)
will meet via diffusive motion in the
characteristic time $\tau \sim \xi^2$,
which explains why the critical dynamic index is $z=2$.
The situation is similar if one is close to (but not at) $T_c=0$.

\begin{figure}
  \begin{center}
        \leavevmode
        \epsfysize=8truecm \epsfbox{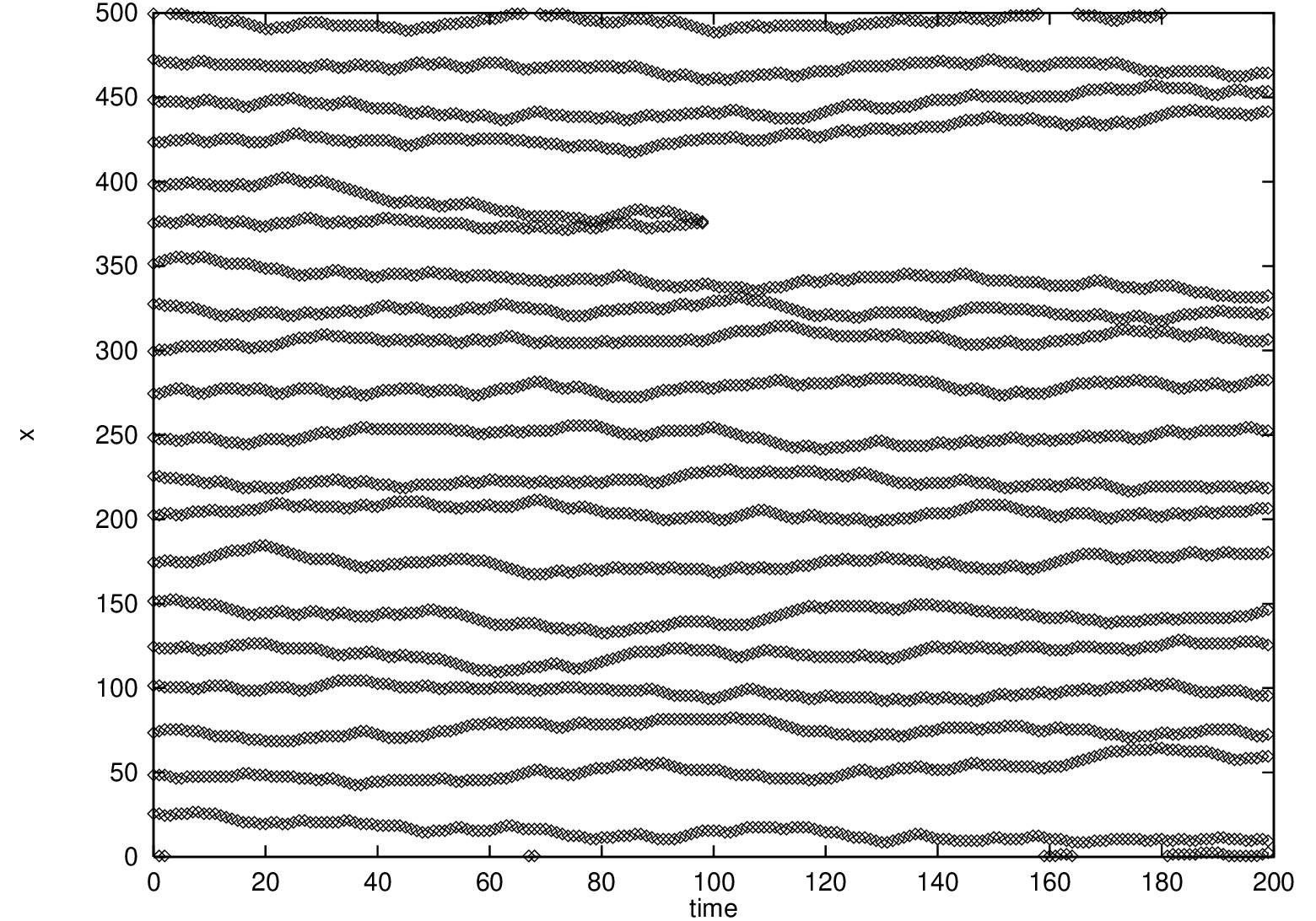}
  \end{center}
  \begin{center}
        \leavevmode
        \epsfysize=8truecm \epsfbox{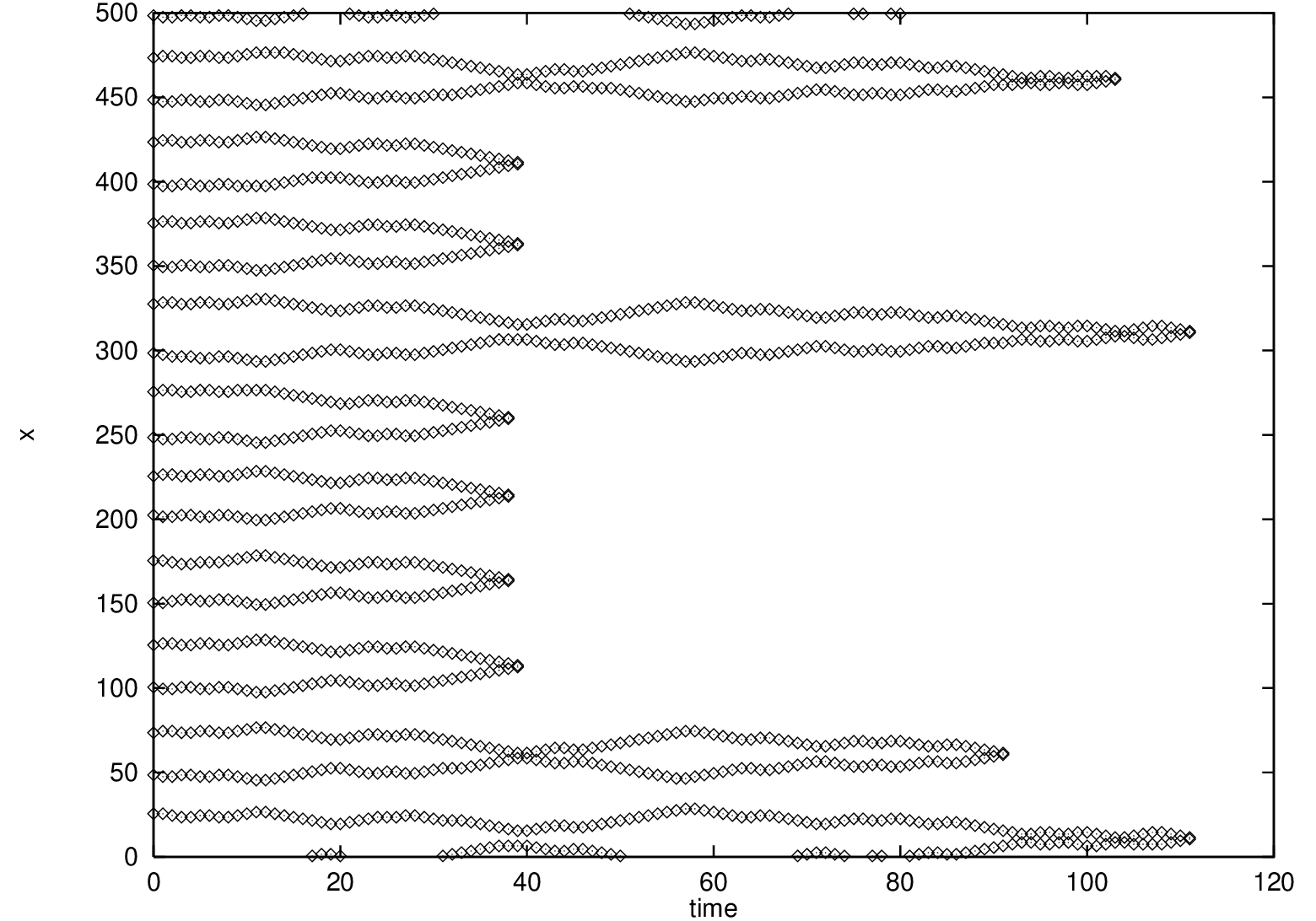}
  \end{center}
        \caption{\it Diffusion of walls for $\kappa=0$ (top) and $\kappa=0.2$ (bottom),
          respectively. Note the different time scales.}
        \label{walls}
\end{figure}

How does this picture change if we switch on the random external field ($\kappa > 0$)? 
The domains parallel
to the external field start growing - both ends of such a cluster will move outwards. Once the
field changes sign, these domains shrink again and the clusters of oppositely aligned spins
grow.
During a longer period of time
these effects compensate each other and the surviving kinks perform effectively a random
walk. However, the annihilation rate of kink-pairs is highly increased. 
Assume, for
example, that after $2T$ iteration steps
the external field had the value $+H_0$ $(T + n)$-times. 
The probability
for this to happen is given by the Bernoulli distribution, 
$B_n^{2T} = {2T \choose n} ({1\over 2})^n$. During this
time, a down-oriented domain whose original length was $L_0$ 
shrinks on average by $2n(2p-1)$,
where both kinks associated with the ends of the domain have moved inward during each of the
$n$ time steps 
with probability $p>1/2$ and outward with probability $q=1-p$. 
Hence, if $2n(2p-1) \sim L_0$, the domain will be eliminated. $p$ is a function
of $\kappa = {H_0\over 2K}$, {\it e.g.} $p(\kappa = 0.2) = 0.982$.
Due to the depletion of small clusters, 
the number of kinks decreases much faster than in the absence of the
external field.
This effect is illustrated by a numerical simulation
in Fig. \ref{walls}, showing the
$T=0$ dynamics of walls for $\kappa = 0$ and $\kappa = 0.2$, respectively. 

Once only a few large clusters remain, however, their width becomes macroscopic.
On this scale the kinks perform again a random walk
and asymptotically one regains $z=2$. 

At finite temperatures, however, the presence of the field term facilitates
the nucleation of new clusters, so that the correlation length \eqref{xi1d} 
(the mean cluster size) is less divergent when $H_0>0$. 

\subsection{The magnetization distribution}

Consider again the map \eqref{mibimap}. In the translational invariant sector
one has
\begin{equation}
        \label{mbimap}
    m(t+1) = \left\{ \begin{array}{lcl} 
                     a + \tilde \gamma m(t) + b c_2(t)\ & {\rm with\ prob.} &\ {1\over 2} \\
                     & & \\
                    -a + \tilde \gamma m(t) - b c_2(t)\ & {\rm with\ prob.} &\ {1\over 2} \\
                        \end{array} \right.
\end{equation}

\begin{figure}
  \begin{center}
        \leavevmode
        \epsfysize=6truecm \epsfbox{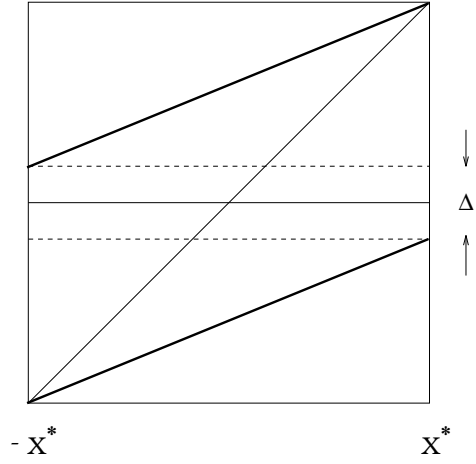}
  \end{center}
        \caption{\it The truncated 1D magnetization map for $\Delta>0$}
        \label{figmap1d}
\end{figure}

As already discussed, the magnetization couples to the correlation function $c_2(t)$, etc.
A simple approximation to decouple the magnetization sector is using for 
$c_2(t)$ the stationary value $c_2 = \eta^2$. 
The resulting map corresponds to a Bernoulli-shift \cite{gh} and
is shown graphically in Fig. \ref{figmap1d}. 
If the gap between the two branches is positive $\Delta>0$

\begin{equation}
\label {1dgap}
\Delta = {2(a+bc_2)(1-2\tilde \gamma)\over 1 -\tilde \gamma} > 0
\end{equation}
the corresponding stationary magnetization distribution is a Cantor set. 

Since $a$, $b$, $(1-\tilde \gamma)$ are positive and $c_2>0$ for ferromagnetic
interactions, the demarcation line between a fractal and 
a nonfractal magnetization distribution
is given by
\begin{equation}
        \label{1dfractalline}
        \tilde \gamma = { 1 \over 2},
\end{equation}
{\it independently} of the actual value $c_2(t)$ might have. 
The time dependence of $c_2(t)$ induces 
nonlinearities in the map. Therefore, although the
distribution remains fractal for $\tilde \gamma < 1/2$, 
in general it is not a homogeneous Cantor set.


\section{Monte Carlo simulations in two dimensions}
\label{sec:mcs}
%

We simulated the RDIM on a two dimensional square lattice on the
neurocomputer SYNAPSE-1/N110. In the following sections we
first describe a Monte Carlo Algorithm (MCA) designed to
make use of the computational power of SYNAPSE-1 and then present
our numerical results. They include a phase diagram in the H-K-plane,
magnetization distributions in the para- and ferromagnetic regime, and
a series of snapshots documenting the behavior of the system.

\subsection{The algorithm}
SYNAPSE-1 is a workstation-driven coprocessor consisting of a systolic
array of eight MA16 Neural Signal Processors. It was kindly put at our
disposal by the ZFE of Siemens AG. Its hardware was designed to
tackle typical problems encountered when simulating neural networks,
namely calculations involving very large matrices. In order to
efficiently make use of the C++-library interface supplied with
SYNAPSE-1 for the RDIM, we devised a MCA that simulates multiples of
eight lattices in parallel.



Consider a square lattice of spins $s^k_{ij} \in \{-1,1\}$ of linear
dimension $L$, where $k$ numbers the system and $i,j=1 \dots L$
denotes the lattice position.  Each of eight lattices is represented
as a column vector $s_{\nu}$ by renumbering indices $\nu = iL + j$.
The eight systems can thus be treated as {\it one} $8 \times
L^2$-matrix.  By setting $\nu \mapsto \nu + L^2$ for $\nu \leq 0$ and
$\nu \mapsto \nu - L^2$ for $\nu > L^2$ we enforce helical boundary
conditions. The neighbors of spin $\nu $ are $\mu = \nu \pm 1, \mu =
\nu \pm L$.

In order to avoid metastable states induced by a simultaneous update
of neighboring spins we split the lattices into black and white sites
in a checkerboard fashion, leading to {\it two} matrices encoding the
eight systems. Note that when a lattice is divided up in this fashion,
if $L$ is odd, the sites in the first and last row have neighbors of
their own color. If $L$ is even, the same is true for the first and
last column of the lattice. For technical reasons we chose $L$ to be
odd.

To further simplify the updating scheme, a copy of the first and last
$L$ components of each lattice are included at the end respectively
beginning of each column vector. A Monte Carlo step now consists of a
parallel update of all black sites followed by an update of all white
sites (or vice versa).

From the well known Glauber dynamic rule Eq.~\eqref{trprob}, a spin is
flipped under the following condition. Given a random number $z \in
[0,1]$ drawn from a uniform distribution, spin $s_{\nu}$ is updated
according to
\begin{equation}
\label{mcflip}
s_{\nu} \mapsto \left\{ 
  \begin{array}{lcl}
    -s_{\nu} & {\rm if } & z < {1 \over 2} ( 1 - \tanh(s_{\nu} (K\Sigma_{\mu} s_{\mu} + H))) \\
    & & \\
    s_{\nu} & {\rm if } & z \geq {1 \over 2} ( 1 - \tanh(s_{\nu} (K\Sigma_{\mu} s_{\mu} + H))) \\
  \end{array}
  \right.
\end{equation}
Usually, spins are either treated sequentially or chosen randomly.
SYNAPSE-1 permits the parallel generation of (pseudo-) random numbers
in a single Elementary Operation (ELOP) which can be piped through a
function lookup table at no extra computational cost. For this reason
we transform the flip condition Eq.~\eqref{mcflip} into
\begin{equation}
\label{synflip}
s_{\nu} \mapsto \left\{ 
  \begin{array}{lcl}
    -s_{\nu} & {\rm if } & \Sigma_{\mu} s_{\nu} s_{\mu} < {1 \over 2K} \log( { {{1 \over 2} - z'} 
      \over { {1 \over 2} + z' } } ) - s_{\nu} { H \over K} \\
    & & \\
    s_{\nu} & {\rm if } & \Sigma_{\mu} s_{\nu} s_{\mu} \geq {1 \over 2K} \log( { {{1 \over 2} - z'} 
      \over { {1 \over 2} + z' } } ) - s_{\nu} { H \over K} \\
  \end{array}
\right.
\end{equation}
where $z'$ is drawn from a uniform distribution in $[-{1 \over 2},{1
  \over 2}]$. The RHS of the flip condition is evaluated in two ELOPs:
One to generate a matrix of random numbers piped through the function
$z' \mapsto {1 \over 2K} \log( { {{1 \over 2} - z'} \over { {1 \over
      2} + z' } } )$ and one weighted matrix subtraction. The LHS also
takes two ELOPs to calculate from the ``black'' and ``white''
matrices. Two further ELOPs are required to construct and evaluate a
flip indicator matrix.  Some more operations are necessary to fix
boundary conditions and to evaluate the mean lattice magnetizations.
This procedure is applied sequentially first to the matrix holding the
``black'' spins and then to the ``white'' one to accomplish a complete
Monte Carlo step.

Initially, the spins in a lattice are set to $+1$ with probability $p$
and $-1$ with probability $1 - p$, where different values of $p$ can
be used for each lattice in one simulation run.  The results for
systems of linear dimension $L=415$ are initialized with $p = $0, 0.2,
0.4, 0.5, 0.6, 0.8, and 1, in addition, there is one lattice in which
the top half of all spins is set to $+1$ and the bottom half to $-1$.
The external driving field is the same for each system. Simulations at
smaller $L$ consist of 64 systems initialized with $p={1 \over 2}$,
but each with its own driving field trajectory.

Temperature $K$ is measured in units of the critical temperature, $K
\to { K \over K_c }$, where $K_c = {J \over k_B T_c} = {1 \over 2} \ln
(\sqrt{2} + 1) \approx 0.44069$ of the standard two dimensional Ising
model, i.e. $K=1$ corresponds to the critical temperature for $H_0=0$.

\subsection{Dynamics and phase diagram}

\begin{figure}
  \begin{center}
    \leavevmode \epsfysize=3truecm \epsfbox{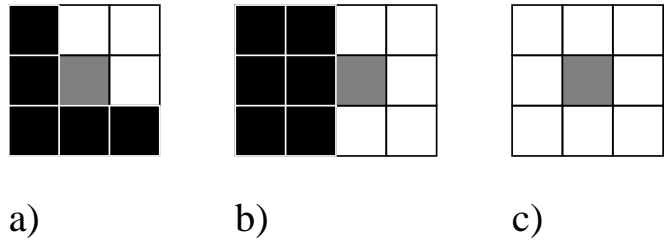}
  \end{center}
        \caption{\it The basic spin-flip configurations, $H_0$ points upwards,
          up (down) spins are black (white), a spin turning up is
          drawn in grey. a) a `droplet'-flip, b) a `domain'-flip, c)
          nucleation-flip}
        \label{flips}
\end{figure}

In order to understand the dynamics of the two-dimensional RDIM in
more detail, it is useful to consider first what happens to a cluster
of parallel spins at $T=0$, in analogy to kink dynamics for the one
dimensional case. Transition rates at $T=0$ are either $0,1$, or
$\frac{1}{2}$.  Recalling that $K$ is measured in units of $K_c$,
define $\kappa':= \frac{H_0}{2K} = \kappa K_c$. Consider now a square
cluster of $2N \times 2N$ parallel spins under the influence of an
anti-parallel external field. First, if $0<\kappa'<1$, the spins at the
corners of the cluster flip with probability $p=1$, all other remain
anti-parallel to the field (as shown in Fig.~\ref{flips} a).  Thus the
cluster disappears if the external field remains constant for $2N-1$
consecutive steps. Secondly, if $1< \kappa' < 2$, such a cluster will
be destroyed in $N$ steps, due to the fact that all but inner spins
will flip with $p=1$. Thirdly, at $\kappa' > 2$ we arrive at a driven
paramagnetic phase. Regardless of their position, all spins will flip
into the direction of the driving field with $p=1$. For the case of
$\kappa' = 0$,$\kappa' = 1$,or $\kappa' = 2$, corner, edge, and inner
spins flip with $p=\frac{1}{2}$.  This implies that, e.g. for a strong
driving field with $\kappa' = 2K_c$, nucleation flips (see
Fig.~\ref{flips} c) may take place {\it inside} the cluster, creating
magnetic swiss cheese.

The transition probabilities of the processes shown in
Fig.~\ref{flips} increase with increasing field strength. At small
fields the a)-type flip is prevalent, resulting in a radial growth
(shrinkage) of clusters. Although energetically more expensive, the
b)-type flip has a large entropy contribution and results in
long-wavelength growth of flat domain walls. The nucleation process
shown in flip c) has the smallest probability.

\begin{figure}
  \begin{center}
    \leavevmode \epsfysize=7truecm \epsfbox{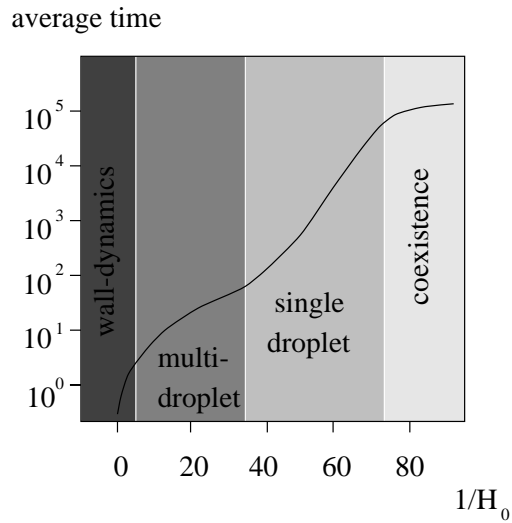}
  \end{center}
        \caption{\it The average time spent in the metastable thermodynamic
          state as a function of inverse field-strength. The different
          domains are denoted according to their main relaxation
          mechanism.  }
        \label{meta}
\end{figure}

\begin{figure}
  \begin{center}
    \leavevmode \epsfysize=7truecm \epsfbox{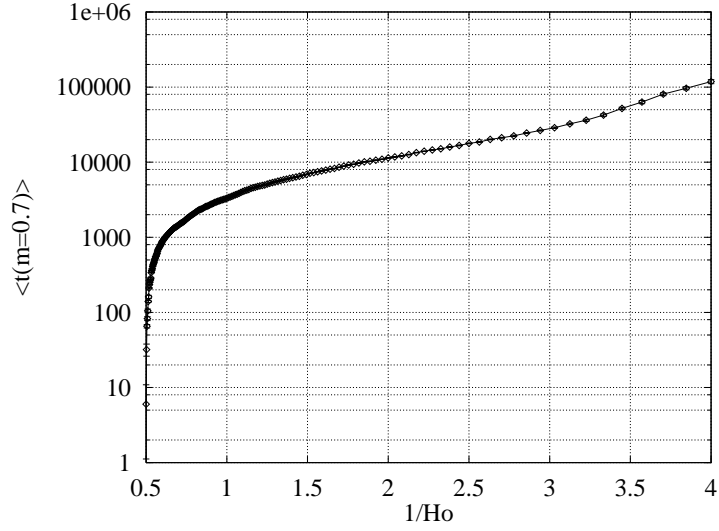}
  \end{center}
        \caption{\it The average first-passage time from $m=-1$ to $m=0.7$ at $K=1.25$ as a function
          of inverse field-strength. $<t(m=0.7)>$ is calculated from
          an ensemble of $64$ systems with linear dimension $L=143$.
          }
        \label{metamc}
\end{figure}

\begin{figure}
  \begin{center}
    \leavevmode \epsfysize=6.5truecm \epsfbox{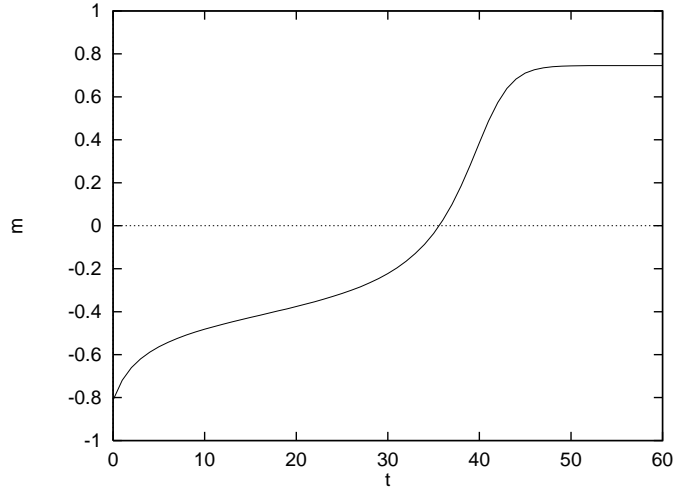}
  \end{center}
        \caption{\it Mean field iteration $m$ vs. time $t$ in an unfavorable field for
          ${H_0 \over K} = 0.057$ and $K=1.2$, which is slightly above
          the critical field.}
        \label{mthighT}
\end{figure}

\begin{figure}
  \begin{center}
    \leavevmode \epsfysize=6.5truecm \epsfbox{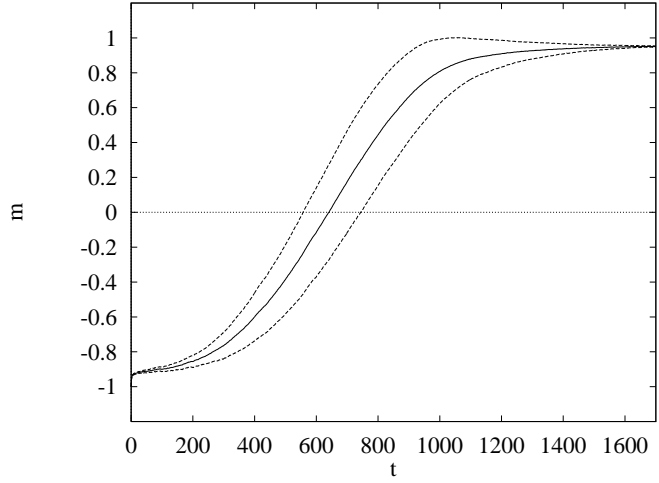}
  \end{center}
        \caption{\it Monte Carlo simulation in an unfavorable field for ${H_0 \over K} = 0.08$,
          $K=1.2$, $\tau_B = 1$, and linear dimension $L=143$.  The
          solid line shows the average magnetization in an ensemble of
          64 systems, $m$, as a function of time $t$. The dashed lines
          mark a 1-$\sigma$ range around the mean.  All systems start
          with all spins down, i.e. $\forall \nu : s_{\nu} = -1$.}
        \label{mcs_uf_highT}
\end{figure}

What happens when switching instantly the field from the equilibrium
state at $-H_0$ into the unfavorable direction $H_0$? The system
relaxes from the now metastable state to the new equilibrium value.
Obviously, the lifetime of the metastable state depends on the
strength of the applied field. This scenario has been discussed in
detail in the ferromagnetic phase using droplet theory and Monte Carlo
simulations (see, for example, \cite{th-mc3}). Here, four distinct
field intervals, shown schematically in Fig. \ref{meta} were,
identified in which the lifetimes markedly differ due to different
decay mechanisms. A numerical result for $K=1.25$ is shown in Fig.
\ref{metamc}, where we approximated the metastable lifetimes by
measuring the average first passage times (FPT) from $m=-1$ to $m=0.7$
in Monte Carlo steps. Figs.~\ref{mthighT} and \ref{mcs_uf_highT} show
examples of the time development of the magnetization from the
metastable state to th new equilibrium for the mean field and 2D
model.

\begin{figure}
  \begin{center}
        \leavevmode
        \epsfysize=8.0truecm  \epsfbox{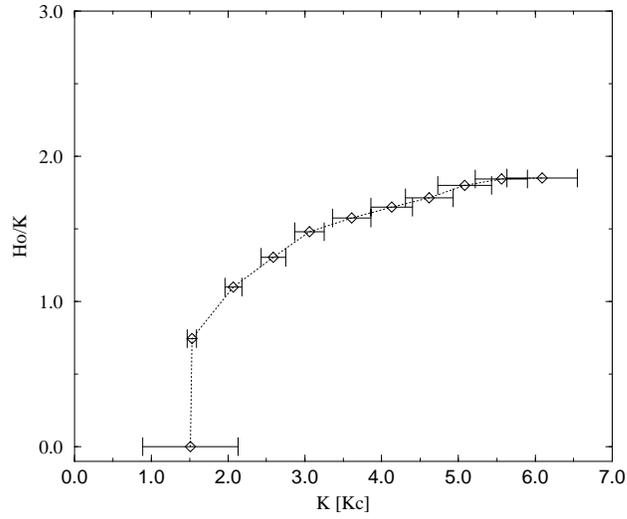}
  \end{center}
  \caption{\it Phase diagram of the 2D RDIM from Monte Carlo simulation of
    64 systems of linear dimension $L=63$.}
        \label{pdHK}
\end{figure}

We calculated numerically a phase diagram in the $K - \frac{H}{K} -
$plane for the 2D RDIM, Fig~\ref{pdHK}. Similar to the mean field and
one dimensional model, there is a paramagnetic, a ferromagnetic (and a
driven paramagnetic) phase. Note that for $K \mapsto \infty$, the
phase boundary should remain below $\kappa' = 2$.  We are currently in
no position to assess this. Also, the behavior around $K=1$ doesn't
seem to correspond to the first order dynamic freezing transition seen
in the mean field theory, rather a second order transition is likely.



\subsection{The paramagnetic phase}

If the external field is above its critical value, $H_0>H_c$, the
stationary phase of the RDIM is paramagnetic. In this phase the system
relaxes relatively fast to the equilibrium state, except close to
$H_c$, where critical slowing down sets in due to the type-I
intermittency effects discussed in the previous section.

This behavior of the average magnetization is shown in
Fig.~\ref{mcs_uf_highT} and is probably enhanced by local correlations
not taken into account in the mean field approximation. If the random
field is switched on, close but above $H_c$ the critical slowing down
is dramatically enhanced.  Further away from the phase transition
point the dynamics is - similarly to the one-dimensional case -
determined by the nucleation and radial growth (shrinking) of
droplet-like clusters.

\begin{figure}
  \begin{center}
    \leavevmode \epsfysize=6.5truecm
    \epsfbox{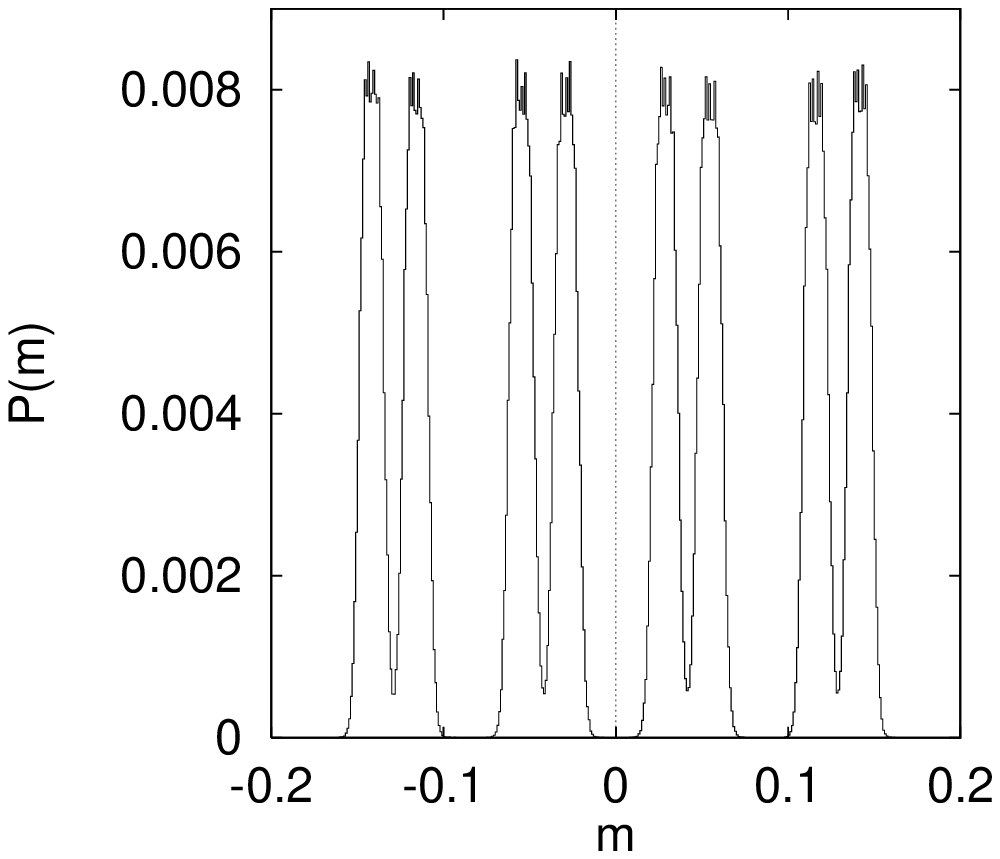}
  \end{center}
  \begin{center}
    \leavevmode \epsfysize=6.5truecm \epsfbox{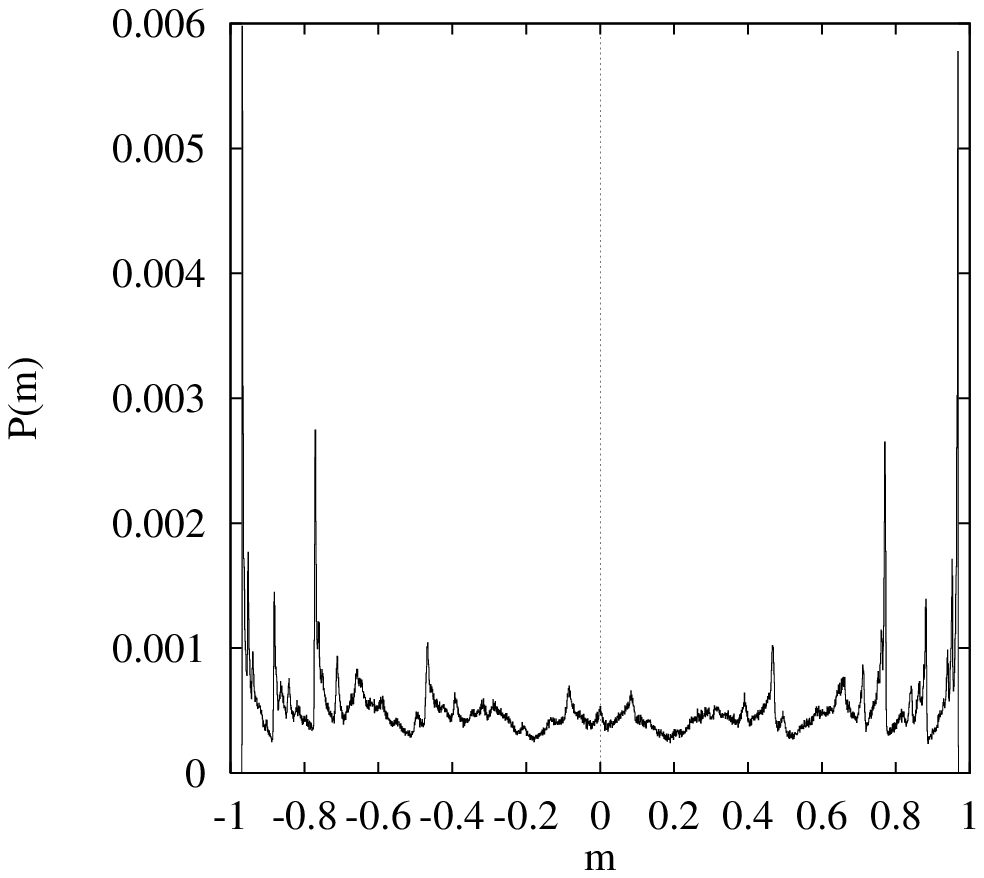}
  \end{center}
  \caption{\it Top: Magnetization distribution for the square lattice RDIM averaged from eight
    different initial conditions for $K=0.4, {H \over K} = 0.5$,
    $\tau_B=1$, and linear dimension $L=415$.  Here, the simution
    covers more than $2 \cdot 10^{5}$ Monte Carlo sweeps. Note the
    similarity to Fig.~1 of Part I.  Bottom: $K=1, {H \over K} = 1$,
    $\tau_B=1$, and $L=415$. The simulation run covers close to $2
    \cdot 10^{5}$ Monte Carlo sweeps. Compare this to
    Fig.~2 of Part I.  }
  \label{2d-mdist-1}
\end{figure}

As expected from the mean-field results, the RDIM can display a
fractal magnetization distribution at higher fields. This is shown in
Fig. \ref{2d-mdist-1}.  Thermal fluctuations and finite size effects
wash out the fine structure of the multifractal magnetization
distribution predicted by the mean-field calculations. However, the
presence of sharp peaks in the distribution (and their scaling
behavior) demonstrates that some of the main features of the
magnetization distribution survive the thermal fluctuations.

These peaks are related to long-lived droplets whose radius is large
enough to allow them to stay alive even when a long series of
unfavorable external field draws makes them shrink. It is, however,
the competition between the two thermodynamically stable states which
leads to a chaotic dynamics and strange attractors.

\subsection{The ferromagnetic phase}

The situation is even more complex below $H_c$. The schematic dependence of
the average time spent in a thermodynamically unstable state {\it vs.} the
inverse of the field strength is shown after \cite{th-mc3} in Fig. \ref{meta}.

\begin{figure}
  \begin{center}
        \leavevmode
        \epsfysize=15truecm \epsfbox{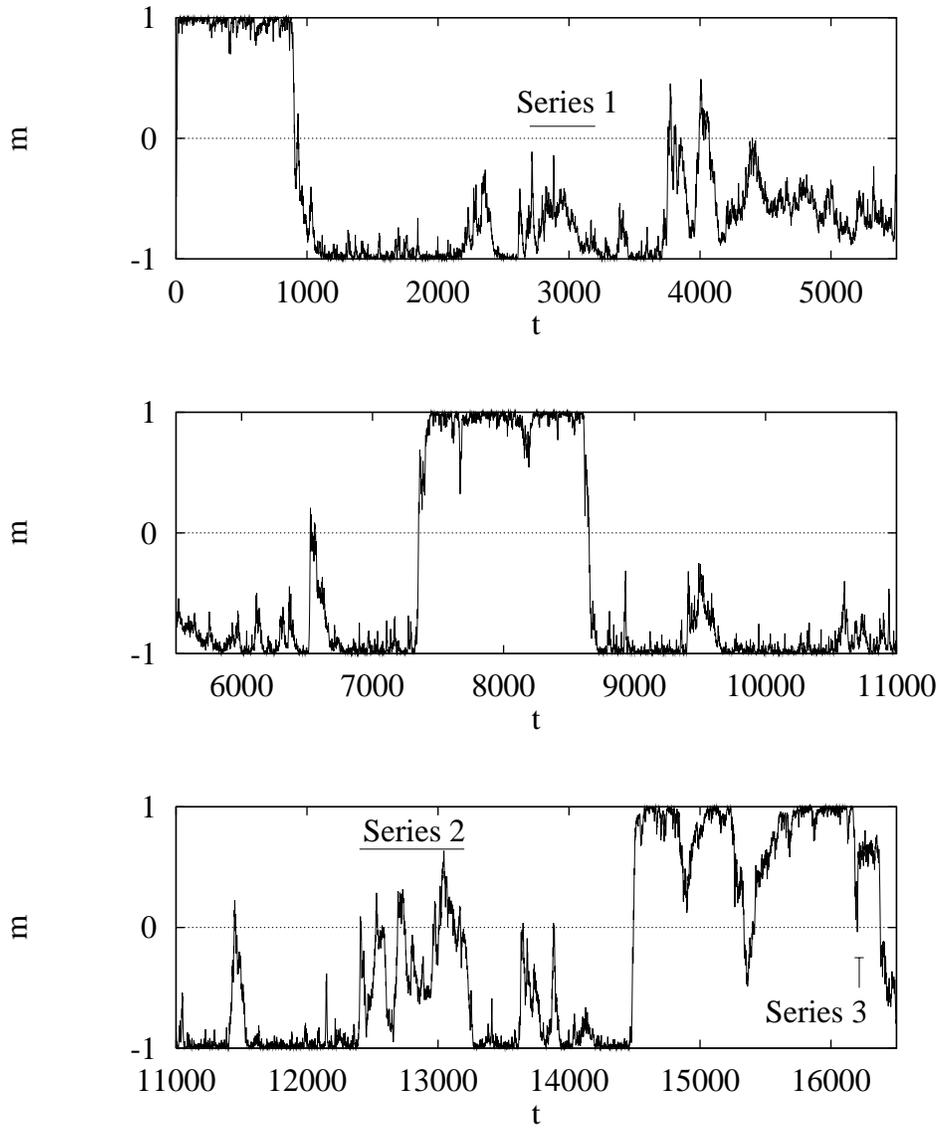}
  \end{center}
        \caption{\it Time series of the mean magnetization of the RDIM. Parameters are 
          $K=2$, ${H \over K} = 1$, $\tau_B=1$, and the linear dimension is $128$. Note that here
          the boundary conditions are periodic. Snapshots of the system are shown in series 1, 2,
          and 3 on the next pages.}
        \label{magserferro}
\end{figure}

The external-field sampling time $\tau_B$ should be chosen in either
the strong or multi-droplet regime (certainly not in the coexistence regime).
By varying $\tau_B$ 
is seems possible to explore these different dynamic mechanisms 
in more detail. 
Fig.~\ref{magserferro}
shows a simulation in the ferromagnetic phase where one can observe both a
multi-droplet (series 1 and 3) and a domain-wall type (series 2) dynamics. 
Again, the {\it  spontaneous magnetization
distribution} shows well separated peaks, which can be seen in Fig.~\ref{2d-mdist-3}.

For the sake of completeness, we show also a Monte Carlo simulation of the 
hysteresis measurement described in Fig.~8 of Part I\cite{part1}. Only the evolution of one
system initialized with spins up/down with equal probability is displayed.
Again, one can see that the thermal fluctuations
are smoothing only the fine scale structure of the mean-field predictions
-- but the main features remain intact.

The results presented here leave open many questions regarding the 2D RDIM -- an
in-depth study by Monte Carlo simulation lies currently beyond our means. The theory
and evolution of droplets and domains in the RDIM as well as hysteretic effects remain
interesting research topics.


\section{Summary and Discussion}
\label{sec:sum}
In Part I and in this article we have discussed in detail
the properties of a simple Ising model subject to a fast switching
external field. In many ways, the situation is just the opposite as in
quenched random systems. While there the defects and hence the
(local) fields are frozen relative to the spin degrees of freedom,
which are (in principle) free to relax, in the RDIM the external field is
the fast variable compared to the interacting spin system. 
From a ``technological''
theoretical point of view the situation is, however, much easier.
We expect that similar analytic results can be obtained
for other random distributions as well. 
Strongly
driven systems show rather peculiar properties, which one could use
for increasing the storage properties of ferromagnetic materials.
For example, it seems possible to use arithmetic coding
in `preparing' a semi-macroscopic ferromagnetic region to fall into a given
distribution peak, as the ones shown in Fig. \ref{2d-mdist-1}. An appropriately
sensitive reading head could then discriminate between the different
magnetization values. Such `devices' could be tested first with the
help of Monte Carlo simulations.
However, the main application domain for randomly driven systems
might well be in biology. Further work in that direction is under progress.



\section*{Acknowledgments}
We are grateful to the ZFE, Siemens
AG and U. Ramacher for the SYNAPSE-1 neurocomputer, on which the Monte
Carlo simulations were performed.  This work was partly supported by
the DFG through SFB 517.

\begin{figure}
  \begin{center}
        \leavevmode
        \epsfysize=17truecm  \epsfbox{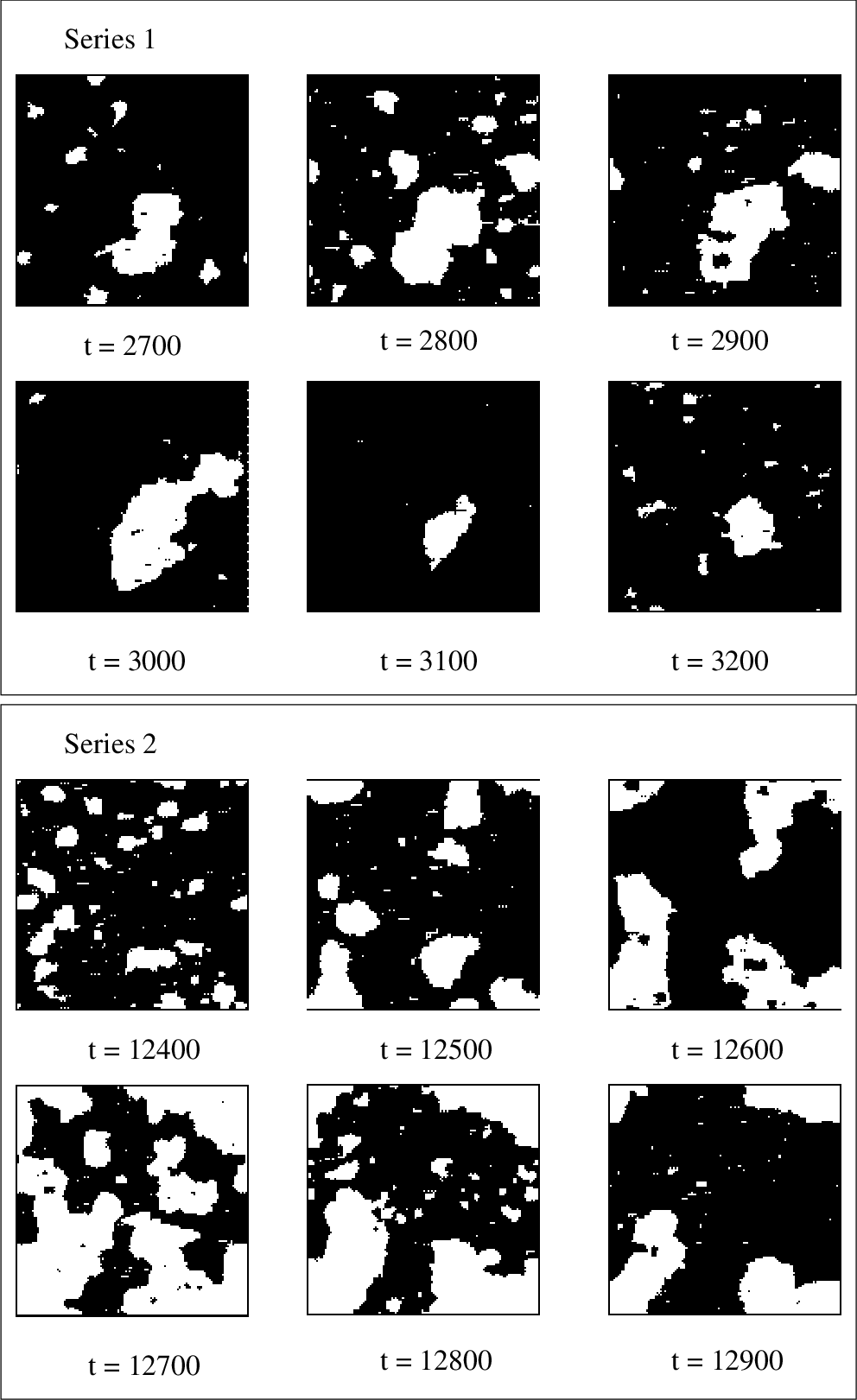}
  \end{center}
        \label{ferrosnap1}
\end{figure}
\begin{figure}
  \begin{center}
        \leavevmode
        \epsfysize=17truecm  \epsfbox{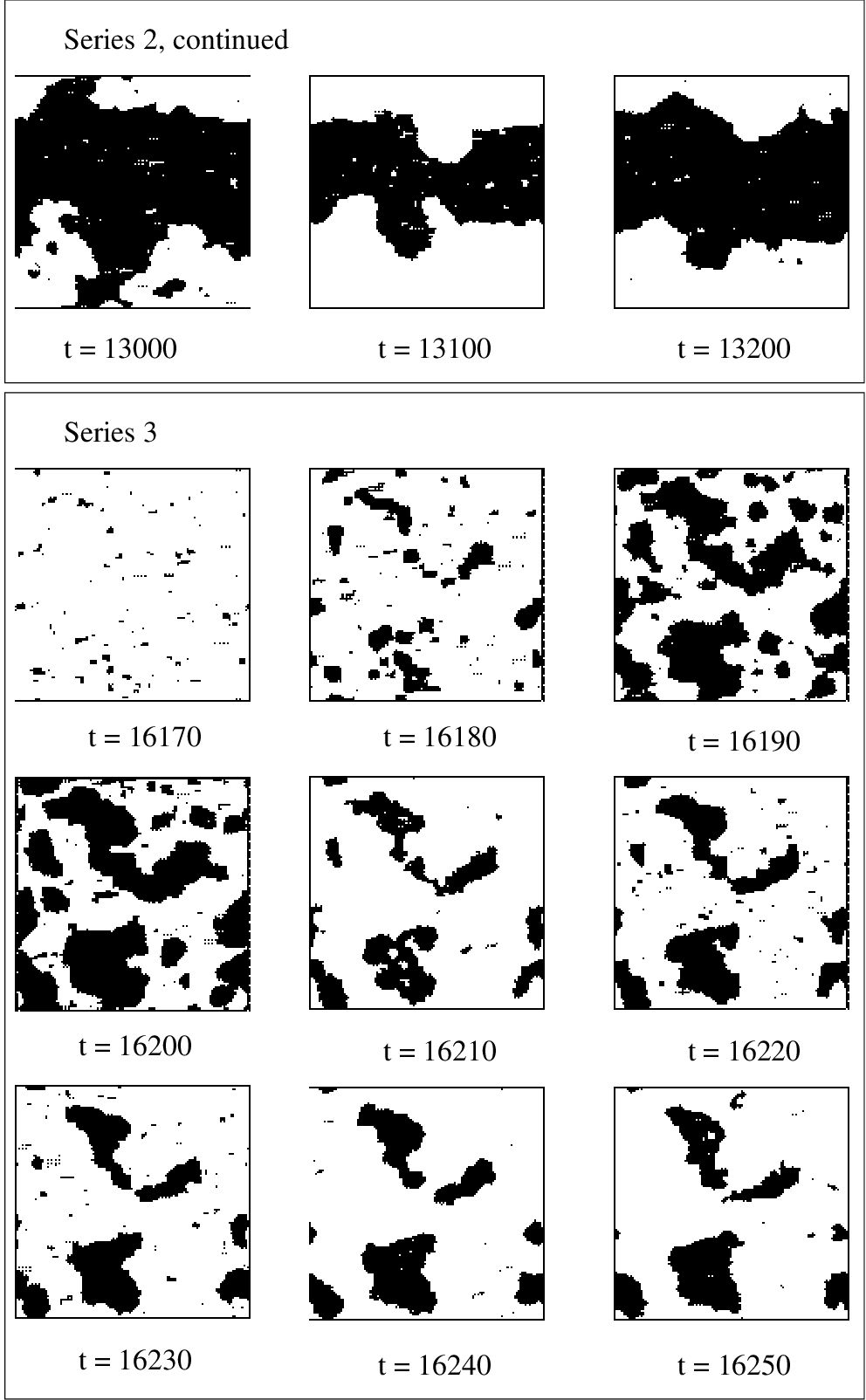}
  \end{center}
        \label{ferrosnap2}
\end{figure}
\begin{figure}
  \begin{center}
    \leavevmode
    \epsfysize=6.5truecm \epsfbox{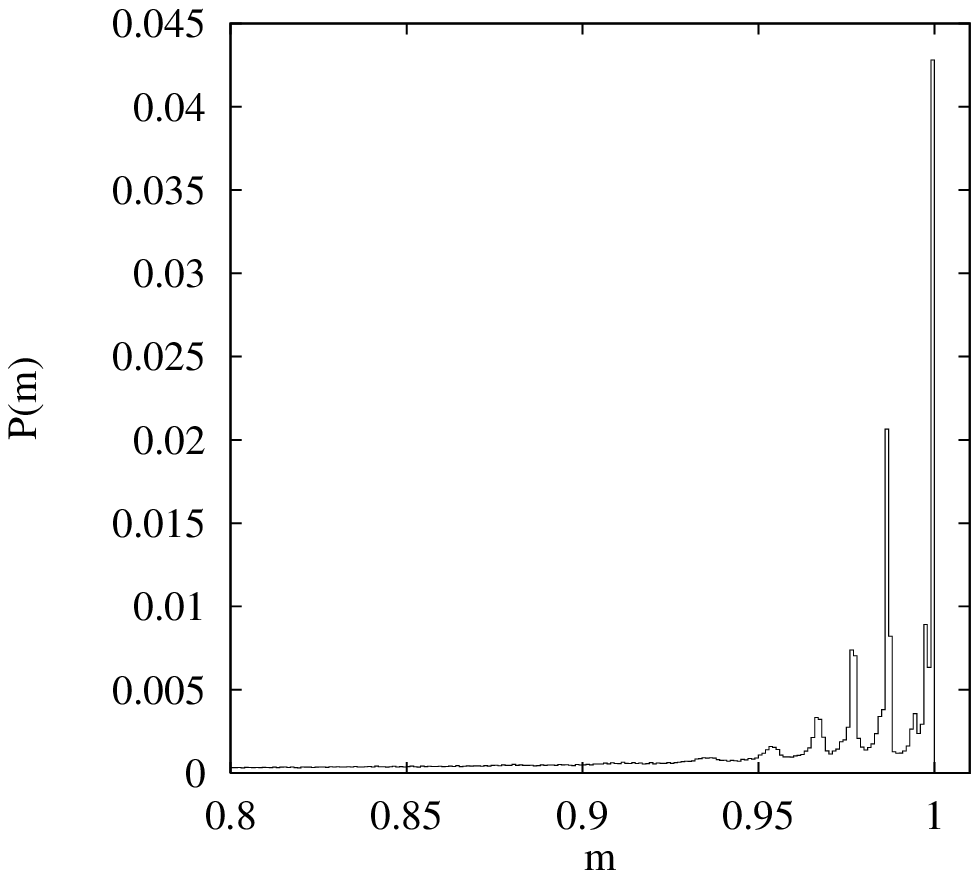}
  \end{center}
  \begin{center}
    \leavevmode
    \epsfysize=6.5truecm \epsfbox{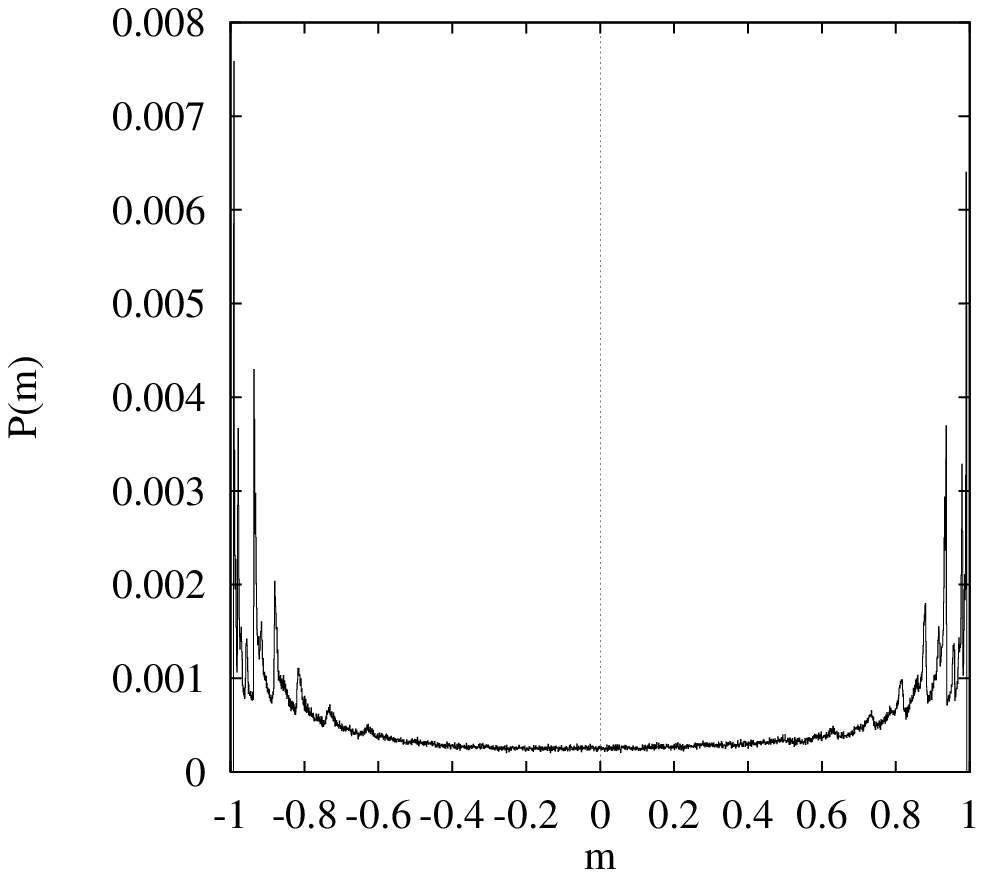}
  \end{center}
  \caption{\it Top: Magnetization distribution for the square lattice RDIM averaged from eight
    different initial conditions for $K=2, {H \over K} = 1$, $\tau_B=1$, and linear dimension $L=415$.
    We simulated more 
    than $2 \cdot 10^{5}$ sweeps. Only the region from $m=0.8$ to $m=1.0$ is shown, the distribution
    is symmetric in $m$. It is similar to the mean field distribution around
    the critical driving field $H_c$ (c.f. Fig.~3 and 4 of Part I).
    Bottom: $K=1.33, {H \over K} = 0.75$, $\tau_B=1$, and $L=415$. We tracked $1.5 \cdot 10^{5}$ Monte 
    Carlo sweeps. Here, the distribution resembles the mean field distribution close to the critical 
    driving field $H_c$.    }
  \label{2d-mdist-3}
\end{figure}
\begin{figure}
  \begin{center}
        \leavevmode
        \epsfysize=6.5truecm \epsfbox{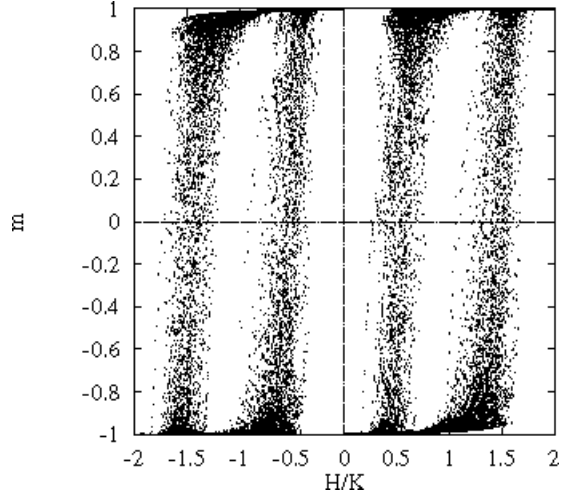}
  \end{center}
        \caption{\it Hysteresis in the square lattice RDIM, showing the mean magnetization $m(t)$ vs.
          the external driving field $H(t)$, see Eq. (32) of Part I. Parameters are 
          $ {A \over K} = {H_0  \over K} =1$, $K=2$, $\Omega = { 2 \pi \over 1000 }$, and linear
          dimension $L=415$. The simulation ran for more than $10^5$ sweeps. 
          See Fig.~8 of Part I for comparison.}
        \label{synhyst}
\end{figure}



\end{document}